\documentclass[11pt]{article}
\usepackage[margin=1in]{geometry}
\usepackage[T1]{fontenc}
\usepackage[utf8]{inputenc}
\usepackage{amsmath,amssymb}
\usepackage{graphicx}
\usepackage{booktabs}
\usepackage{url}
\usepackage{xcolor}
\usepackage[hidelinks]{hyperref}
\usepackage{caption}
\newcommand{\authorname}{Ekkachai Jueng}
\newcommand{\authoremail}{ekkachai@sci.rmuti.ac.th}

\title{\textbf{Q-Learning Lab: Teaching Reinforcement Learning\\Through Learner-Generated Trace Analysis}}
\author{\authorname\\
Computer Science Program\\
Faculty of Sciences and Liberal Arts\\
Rajamangala University of Technology Isan\\
Nakhon Ratchasima, Thailand\\
\texttt{\authoremail}}
\date{July 2026}

\begin{document}
\maketitle

\begin{abstract}
Reinforcement learning is usually introduced through the Bellman update, yet the equation often remains abstract to undergraduates: they watch policy arrows converge but rarely observe \emph{how} each value is computed or \emph{why} an action is chosen. We present \textbf{Q-Learning Lab}, a single-file, browser-based, bilingual (Thai/English) tool for teaching tabular Q-learning that requires no installation and no network connection. Beyond the usual gridworld visualization---color-coded Q-values and policy arrows on a $5\times5$ world---the tool exposes a live Bellman-substitution panel that shows the numeric update at every step, and logs each transition, including the full pre-action Q-row, the greedy-versus-random decision under $\varepsilon$-greedy exploration, and wall-collision events, into an exportable trace. The central contribution is a \emph{learn--export--analyze} loop: learners run their own agent, export the complete trace as CSV, and then analyze it themselves---producing learning curves, value heatmaps, and visitation maps---turning a passive demonstration into a source of learner-generated data for reflective inquiry. We validate the tool without human-subject data through three complementary evaluations: (i) correctness of the learned values and policy against a value-iteration ground truth on the identical MDP; (ii) hyperparameter sweeps ($\alpha$, $\gamma$, $\varepsilon$) demonstrating that every pedagogical claim the tool makes is reproducible; and (iii) a reward-editing study that uses the ground-truth optimal policy to separate two behaviorally identical but diagnostically opposite failure modes---an exploration failure versus genuine reward misspecification---that a single edited reward can produce. We also situate the tool against existing gridworld visualizers via a feature comparison, describe its grounding in learning-by-doing pedagogy, and include a 50-minute lesson plan. The tool and all experiment code are openly available.
\end{abstract}

\section{Introduction}

Reinforcement learning (RL) has moved from a specialist topic to core curriculum: it underlies game-playing agents, robotics, recommendation, and---through reinforcement learning from human feedback---the alignment of large language models. Tabular Q-learning \cite{watkins1992} is the standard entry point, and virtually every introductory course presents the same object: a gridworld, an agent, and the Bellman-style update
\begin{equation}
Q(s,a) \leftarrow Q(s,a) + \alpha\,\bigl[\,r + \gamma \max_{a'} Q(s',a') - Q(s,a)\,\bigr].
\label{eq:qlearning}
\end{equation}

In our experience teaching undergraduate machine-learning courses in Thailand, the difficulty is rarely the gridworld itself. Students readily accept that an agent wanders, collects rewards, and eventually finds a path. The difficulty is the \emph{mechanism}: what exactly is inside the brackets of Eq.~\ref{eq:qlearning} at step 137 of episode 12? Which of the four Q-values did the agent look at before it moved, and did it exploit or explore? Why does the value of a state two cells away from the goal only start rising in the third episode? Existing teaching tools---from the classic Berkeley CS188 gridworld \cite{denero2010} to recent browser-based playgrounds \cite{juliani2025}---answer these questions only indirectly: they display the \emph{outcome} of learning (colored cells, policy arrows) while the update itself stays hidden inside the code.

Q-Learning Lab is built around the opposite premise: every quantity in Eq.~\ref{eq:qlearning} should be \emph{inspectable at every step}, and the record of those steps should belong to the learner. The tool makes three design moves.

\paragraph{Live Bellman substitution.} A dedicated panel re-renders Eq.~\ref{eq:qlearning} after every single step with the actual numbers substituted in: the observed reward $r$, the bootstrap term $\max_{a'}Q(s',a')$, the resulting target, the temporal-difference (TD) error, and the old and new $Q(s,a)$. Terminal transitions and wall collisions are annotated explicitly, so students see, for instance, that a terminal state contributes $\max_{a'}Q(s',a')=0$, and that on a wall bump the successor state equals the current state---so the update bootstraps from the agent's own value, a subtlety that otherwise goes unnoticed.

\paragraph{Decision-complete trace logging.} Every transition is logged with fourteen on-screen fields, and additional fields in the exported file, including the entire Q-row of the current state \emph{before} the action was selected, which action the greedy argmax would have chosen, whether the executed action was greedy or random under $\varepsilon$-greedy, the TD target and error, and the pre-/post-update Q-values. The log is thus not merely a movement history but a complete record of the agent's decision process---a visible, human-readable analogue of a replay buffer.

\paragraph{The learn--export--analyze loop.} The trace can be exported as a CSV file (UTF-8 with BOM, CRLF line endings, so it opens cleanly in spreadsheet software) and re-imported into any analysis environment. Learners therefore do not analyze a canned dataset: they analyze \emph{data their own agent generated minutes earlier}, plotting learning curves, per-state value heatmaps, visitation counts, and the shifting exploration/exploitation ratio across episodes. This closes a reflective loop---act, record, inspect---that a pure animation cannot provide.

The tool is a single HTML file with no external dependencies; it runs offline in any modern browser, can be projected in a lecture, embedded in slides via an \texttt{iframe}, or handed to students as a file. The interface is bilingual, switching between Thai and English with one button, which matters in classrooms where the instructional language is Thai but the field's terminology is English.

The goal of this paper is technical and pedagogical \emph{design} validation rather than measurement of learning gains. Because the tool targets classrooms, one might expect a user study; we deliberately evaluate without human subjects in this first paper. Instead we ask a question that user studies usually skip: \emph{is everything the tool shows and claims actually true of the algorithm it implements?} Section~\ref{sec:eval} verifies the implementation against a value-iteration ground truth, reproduces every hyperparameter lesson in the built-in lesson plan as a multi-seed experiment, and uses the ground-truth optimal policy to correctly diagnose what happens when rewards are edited. Classroom evaluation is future work (Section~\ref{sec:limitations}).

\section{Related Work}
\label{sec:related}

Gridworld visualizations of value-based RL are as old as the pedagogy itself. The UC Berkeley CS188 project suite \cite{denero2010} remains the most influential: students implement value iteration and Q-learning against a gridworld with graphical overlays of values and Q-values, and can drive the agent manually to set up specific scenarios. It is, however, a Python codebase that students must install and partially implement---appropriate for a programming-heavy AI course, but a high barrier for a conceptual first encounter or for classrooms with heterogeneous computing environments.

More recent tools lower the barrier by moving to the browser. The Beginner's RL Playground \cite{juliani2025} runs tabular methods interactively on the web with adjustable algorithms and exploration strategies. The Virtual Labs Q-learning experiment \cite{virtuallabs} lets students watch Q-values update in a gridworld after each action. Smaller open-source projects such as \texttt{tinyrl} \cite{tinyrl} serve value iteration and Q-learning visualizations from a local web server with tunable $\alpha$ and $\varepsilon$, and \texttt{gridworld-qlearning} \cite{rasoul} animates live training through an interactive Matplotlib viewer.

All of these tools visualize the \emph{state} of learning. Among the representative tools reviewed here, we did not find a system that simultaneously (i) renders the update equation itself with live numeric substitution at every step, (ii) logs the agent's full pre-decision Q-row together with the greedy/random provenance of each action, and (iii) exports the complete decision trace for learner-side analysis; Table~\ref{tab:features} summarizes the comparison. We make no claim of exhaustiveness over the many RL demonstrations available online, but across the tools commonly cited for teaching, the gap is consistent: existing tools treat the learner as a spectator of a converging process, whereas Q-Learning Lab treats the learner as an analyst of a recorded process. Our contribution is not a better animation but a different epistemic relationship between the learner and the algorithm.

The pedagogical stance is constructionist-inspired and draws on two well-established traditions. Constructionism \cite{papert1980} holds that learning is most effective when learners build and inspect artifacts of their own making; here the artifact is the trace dataset the learner's agent produces. Experiential learning theory \cite{kolb1984} describes a cycle of concrete experience, reflective observation, and abstract conceptualization; the learn--export--analyze loop is a direct operationalization of that cycle, with the CSV trace serving as the object of reflective observation. Karpathy's single-file, dependency-free implementations such as \texttt{micrograd} \cite{karpathy2020} motivated the engineering constraint that the entire system remain one readable file.

The design commitment to \emph{mechanistic transparency} is not unique to introductory settings, and it is worth noting where the same commitment surfaces at the opposite end of the expertise spectrum. The Predator--Prey Archetype Gridworld Environment \cite{atif2025}, a modular multi-agent RL testbed, is motivated by the observation that most MARL environments hide their transition mechanics and entangle environment logic with learning code; it responds by enforcing a strict separation between dynamics, perception, incentives, and learning, by making reward functions an explicit plug-in point so that shaping effects can be studied in isolation, and by treating reproducibility as a constraint enforced through configuration and explicit seeds rather than a property hoped for. Those aims---inspectable dynamics, rewards as an object of experiment, reproducibility by construction---are the same aims that motivate Q-Learning Lab, even though the two systems sit at opposite ends of the audience spectrum: one is research infrastructure for multi-agent experimentation, deployed as a Python package; the other is a single-file browser tool whose target user is meeting the Bellman update for the first time. That the same principle is arrived at independently for a MARL research testbed and for a first-week teaching aid suggests it is a property of good RL tooling generally, not a concession to novices.

Our use of the phrase \emph{learner-generated trace} should also be positioned against learning analytics and educational data mining \cite{siemens2013, baker2009}, where trace data conventionally means logs of \emph{learner} behavior collected at scale and analyzed by researchers or institutions to model, predict, or intervene, and against trace methodologies for studying self-regulated learning \cite{winne2010}. This paper inverts that arrangement: the trace records the behavior of the learner's \emph{agent}, not of the learner, and the analyst is the learner themself rather than an institution. We borrow from these fields the core idea that a fine-grained event trace is a legitimate object of reflective analysis, but the tool is an educational instrument for a single learner, not a learning-analytics system, and we make no analytics claims.

\begin{table}[t]
\centering
\small
\caption{Feature comparison with representative gridworld Q-learning teaching tools. LB = live Bellman substitution panel; QR = per-step logging of the full pre-action Q-row and greedy/random provenance; CSV = one-click export of the complete trace; RW = rewards editable in the interface; 1F = single self-contained file; TH/EN = bilingual Thai/English interface. Entries reflect the publicly documented feature set of each tool at the time of writing.}
\label{tab:features}
\begin{tabular}{lcccccccc}
\toprule
Tool & Browser & No install & LB & QR & CSV & RW & 1F & TH/EN \\
\midrule
CS188 Gridworld \cite{denero2010} & -- & -- & -- & -- & -- & partial & -- & -- \\
RL Playground \cite{juliani2025} & \checkmark & \checkmark & -- & -- & -- & -- & -- & -- \\
Virtual Labs \cite{virtuallabs} & \checkmark & \checkmark & -- & -- & -- & -- & -- & -- \\
\texttt{tinyrl} \cite{tinyrl} & \checkmark & -- & -- & -- & -- & -- & -- & -- \\
\texttt{gridworld-qlearning} \cite{rasoul} & -- & -- & -- & -- & -- & partial & -- & -- \\
\textbf{Q-Learning Lab (ours)} & \checkmark & \checkmark & \checkmark & \checkmark & \checkmark & \checkmark & \checkmark & \checkmark \\
\bottomrule
\end{tabular}
\end{table}

\section{System Design}
\label{sec:system}

\subsection{Environment}

\begin{figure}[t]
\centering
\includegraphics[width=0.42\linewidth]{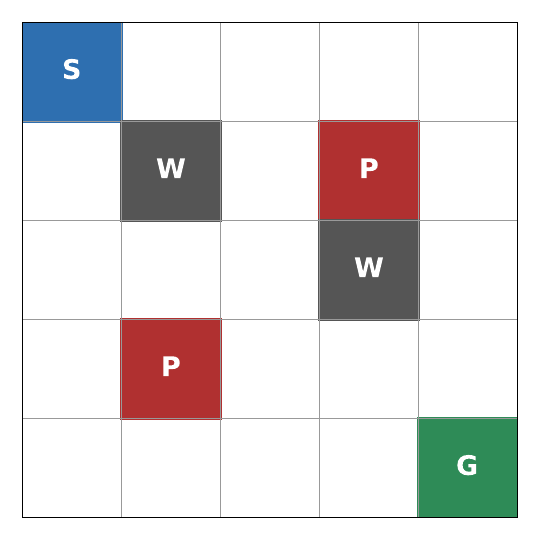}
\caption{The environment. S = start, G = goal ($+10$, terminal), P = pit ($-10$, terminal), W = wall (impassable; bumping costs $-0.1$ and the agent stays in place). Every step costs $-0.1$.}
\label{fig:env}
\end{figure}

The environment (Figure~\ref{fig:env}) is a deterministic $5\times5$ gridworld. The agent starts at the top-left cell $(0,0)$ and must reach the goal at $(4,4)$. Two pit cells at $(1,3)$ and $(3,1)$ terminate an episode with a penalty; two wall cells at $(1,1)$ and $(2,3)$ are impassable. Actions are the four cardinal moves. An action into a wall or off the grid leaves the agent in place (a ``bump''). Default rewards are $+10$ for reaching the goal, $-10$ for falling into a pit, $-0.1$ per step, and $-0.1$ per bump; all four values are editable in the interface, which enables the reward-shaping exercises of Section~\ref{sec:misspec}. The optimal path takes eight steps, for an undiscounted optimal return of $7\times(-0.1)+10 = 9.3$.

The layout is deliberately small enough that the entire Q-table (25 cells $\times$ 4 actions) is visible on screen at once, yet contains the three ingredients that generate the interesting lessons: a hazard adjacent to the direct path, a wall that forces a detour, and enough open space that exploration visibly matters.

\subsection{Interface}

The interface has five coordinated regions. (1) The \emph{grid} renders each cell as four triangles, one per action, colored from red (low Q) through gray (near zero) to green (high Q), with a white arrow showing the current greedy policy and a yellow dot for the agent. (2) The \emph{hyperparameter card} exposes $\alpha$ (default $0.30$), $\gamma$ (default $0.95$), and $\varepsilon$ (default $0.30$) as sliders whose changes take effect immediately, alongside running statistics (episode count, steps, goals reached, last return). (3) The \emph{rewards card} makes all four reward values editable. (4) The \emph{Bellman panel} shows Eq.~\ref{eq:qlearning} for the most recent step with every symbol replaced by its numeric value, decomposed into reward, bootstrap term, target, TD error, and the old-to-new Q movement. (5) The \emph{history table} lists the most recent 500 transitions, newest first, with episode boundaries marked.

Execution modes match distinct teaching moments: a \emph{single-step} button for slow, narrated walkthroughs; a \emph{continuous-play} mode with a speed slider; and a \emph{train $\times$100 episodes} mode that skips logging for speed and is used to demonstrate convergence. Clearing the Q-table and clearing the log are separate actions, so an instructor can, for example, retrain under new rewards while preserving the old trace for comparison.

The interface language toggles between Thai and English with a single button; all labels, hints, the explanation panel, and dynamically generated strings (decision chips, terminal annotations) switch together. The exported CSV uses fixed ASCII column names in both languages, so downstream analysis code is language-independent.

\subsection{The trace and the learn--export--analyze loop}
\label{sec:loop}

Each logged transition records: global step, episode, step-within-episode, state coordinates, the four Q-values of the state \emph{before} action selection, the greedy action, an exploration flag (1 if the action was sampled uniformly under $\varepsilon$-greedy, 0 if it was the argmax), the executed action, reward, successor coordinates, a bump flag, $\max_{a'}Q(s',a')$, the TD target, the TD error, the pre- and post-update $Q(s,a)$, and a terminal flag. The on-screen table renders the same information compactly---the greedy action highlighted, the executed action bracketed---so a student can see at a glance when the agent acted against its own best estimate.

\begin{figure}[t]
\centering
\includegraphics[width=0.495\linewidth]{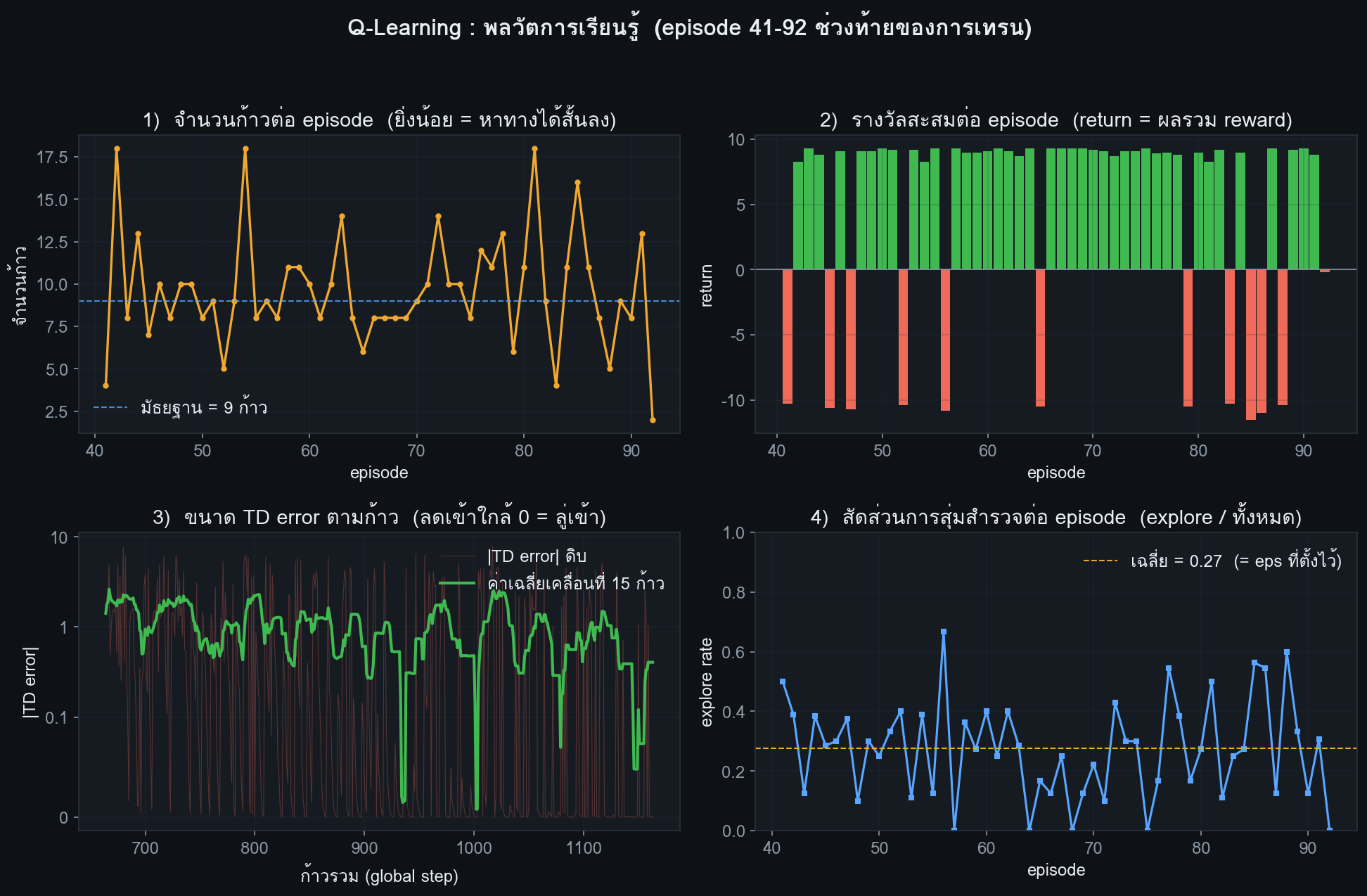}\hfill
\includegraphics[width=0.495\linewidth]{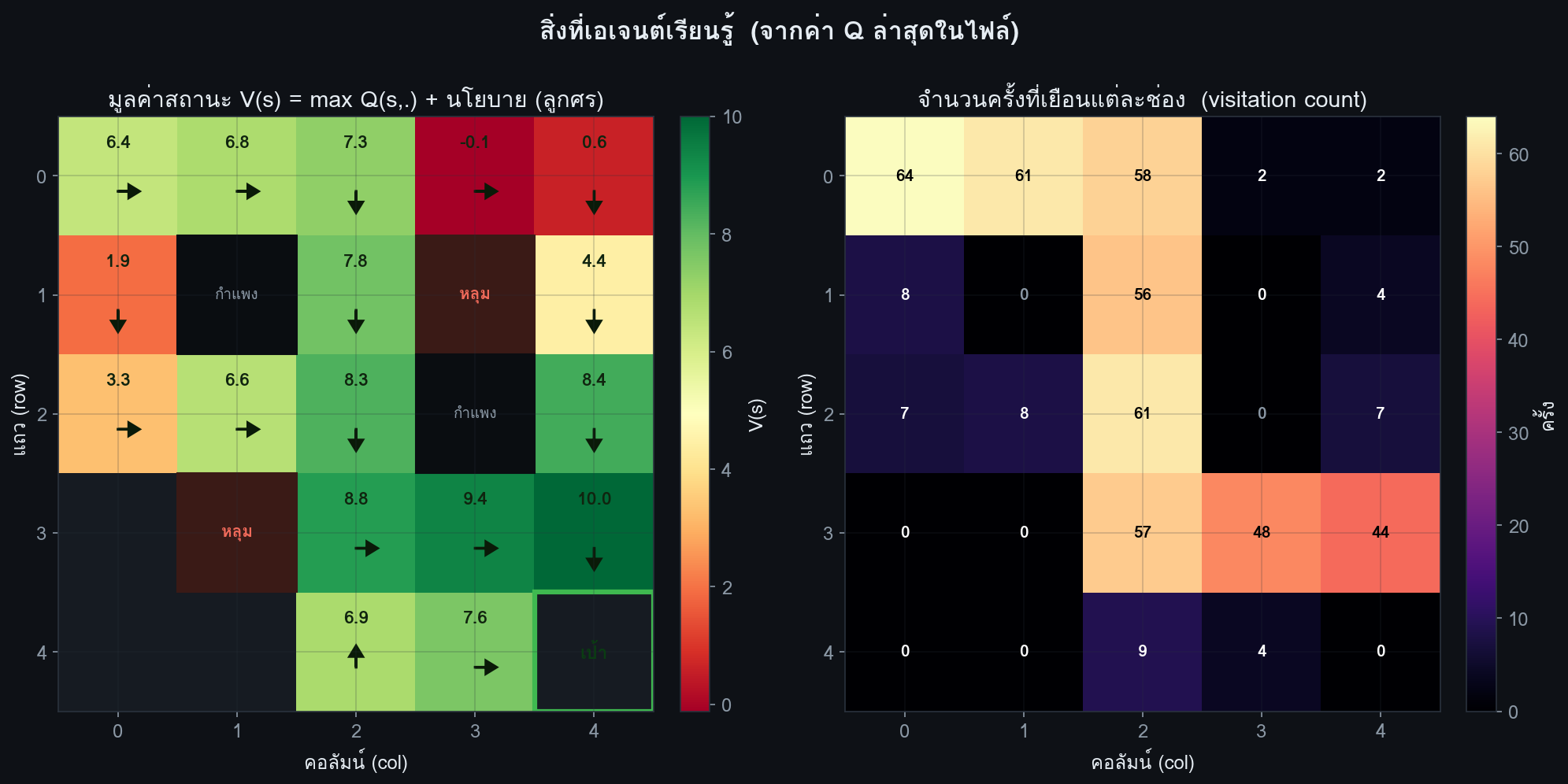}
\caption{Learner-side analysis of a trace exported from the tool (learning dynamics, left; value function $V(s)=\max_a Q(s,a)$ with policy arrows and state-visitation counts, right). These figures were produced from an actual CSV exported by the tool, exactly as a student would in the analysis phase of the lesson.}
\label{fig:trace}
\end{figure}

The one-click CSV export turns this log into a dataset. In the accompanying lesson, students train an agent, export the trace, and answer analysis questions in a spreadsheet or notebook: plot return per episode; compute the exploration ratio per episode; build a visitation heatmap and compare it with the value heatmap; find the first large positive TD error and explain where it occurred. Figure~\ref{fig:trace} shows learning curves and value/visitation heatmaps produced from an actual trace exported from the tool. Because the data is the learner's own---generated under hyperparameters the learner chose---the analysis is an act of reflection on their own experiment rather than an exercise on foreign data. This is the constructionist-inspired core of the design \cite{papert1980, kolb1984}: the trace is an inspectable artifact of the learner's own making, and the analysis step converts a concrete experience into abstract conceptualization.

The history table also serves as a forward pointer: it is, visibly, a replay buffer. When the course later reaches Deep Q-Networks \cite{mnih2015}, students have already spent an hour reading one.

\subsection{Implementation}

The entire system---environment, agent, renderer, logger, and both language packs---is one HTML file (roughly 600 lines of HTML/CSS/JavaScript) with no external dependencies, no build step, and no network access. It runs from a double-click on any modern browser, which removes the installation barrier that limits tools like the CS188 suite in non-CS classrooms, and makes the source short enough to be read end-to-end by a motivated student, in the spirit of single-file pedagogical implementations \cite{karpathy2020}.

\section{Pedagogical Design and Lesson Plan}
\label{sec:pedagogy}

The tool ships with a 50-minute lesson plan structured around five moments, each mapped to a specific interface affordance.

\emph{(1) Stepping through ignorance (10 min).} Using single-step mode on a fresh Q-table, the instructor shows that early TD errors consist of the immediate reward alone, and that a greedy policy over an all-zero table is meaningless---grounding the need for exploration before the word ``exploration'' is introduced.

\emph{(2) The first goal (10 min).} The class steps until the agent first reaches the goal and watches the large positive TD error appear in the Bellman panel; in subsequent episodes the panel shows that value flowing backwards, one cell per visit, from the goal toward the start. In our experience this backward propagation is the single most common point of confusion in Q-learning, and it is precisely what a live substitution panel makes visible.

\emph{(3) Convergence (5 min).} The $\times$100 training mode is pressed a few times; policy arrows align into a path and the green gradient spreads across the grid.

\emph{(4) Hyperparameter experiments (15 min).} Students set $\gamma=0$ (myopia: only goal-adjacent cells turn green), $\gamma=0.95$ (value flows everywhere), $\varepsilon\in\{0,1\}$ (pure exploitation locks in; pure exploration never focuses), and $\alpha\in\{0.05,0.9\}$ (slow-but-stable versus fast-but-volatile updates). Section~\ref{sec:sweeps} verifies each of these claims experimentally.

\emph{(5) Reward design (10 min).} Students set the step penalty to $0$ (the agent stops hurrying) and raise the wall penalty (the agent learns to avoid walls sooner). They then set the pit reward to $+5$ and watch the agent dive into pits---and, crucially, check that behavior against the optimal policy, discovering that the agent is merely under-explored rather than correctly serving a broken objective. Raising the pit reward further ($+20$) produces a genuinely pit-seeking optimal policy. Section~\ref{sec:misspec} treats these two cases in full; together they teach that diagnosing an RL agent requires comparing its behavior with the Bellman-optimal policy, not reading the behavior alone \cite{amodei2016}.

The lesson closes with discussion questions that the trace makes answerable from evidence rather than authority: Why did the Q-values of wall-bumping actions decay without any explicit instruction? What does the update bootstrap from when $s'=s$? Would $\varepsilon$-decay beat a constant $\varepsilon$ here?

\section{Evaluation}
\label{sec:eval}

We evaluate without human-subject data, using three questions: Is the algorithm the tool implements \emph{correct}? Are the lessons the tool teaches \emph{reproducible}? And when a student edits the rewards, does the tool support a \emph{correct diagnosis} of the agent's behavior? All experiments run on a Python reference implementation that mirrors the JavaScript environment constant-for-constant (grid layout, rewards, bump semantics, $\varepsilon$-greedy tie-breaking by argmax); the reference implementation and every experiment script are released with the tool. Unless stated otherwise, results aggregate 10 random seeds.

\subsection{Correctness against value iteration}
\label{sec:correctness}

Because the MDP is small and deterministic, exact ground truth is available: value iteration to numerical convergence (tolerance $10^{-12}$) yields $V^\ast$ and the set of optimal actions per state; the optimal value of the start state under $\gamma=0.95$ is $V^\ast(s_0)=6.380$. We then run tabular Q-learning exactly as the tool defines it, with $\varepsilon=0.3$ behavior policy and a Robbins--Monro learning-rate schedule $\alpha_n = n(s,a)^{-0.75}$ for 50{,}000 episodes.

Gridworlds routinely admit several equally optimal actions per state, so naive argmax-to-argmax comparison would understate agreement. We therefore score a state as correct when the learned greedy action lies in the optimal action set:
\begin{equation}
\mathrm{agreement} \;=\; \frac{1}{|\mathcal{S}^{\circ}|}\sum_{s\in\mathcal{S}^{\circ}} \mathbb{1}\!\left[\arg\max_a Q(s,a) \in A^\ast(s)\right],
\qquad
A^\ast(s) = \Bigl\{a : r(s,a) + \gamma V^\ast(s') = V^\ast(s)\Bigr\},
\label{eq:agreement}
\end{equation}
where $\mathcal{S}^{\circ}$ is the set of non-terminal states and ties in $A^\ast(s)$ are identified numerically with tolerance $10^{-9}$.

\begin{figure}[t]
\centering
\includegraphics[width=\linewidth]{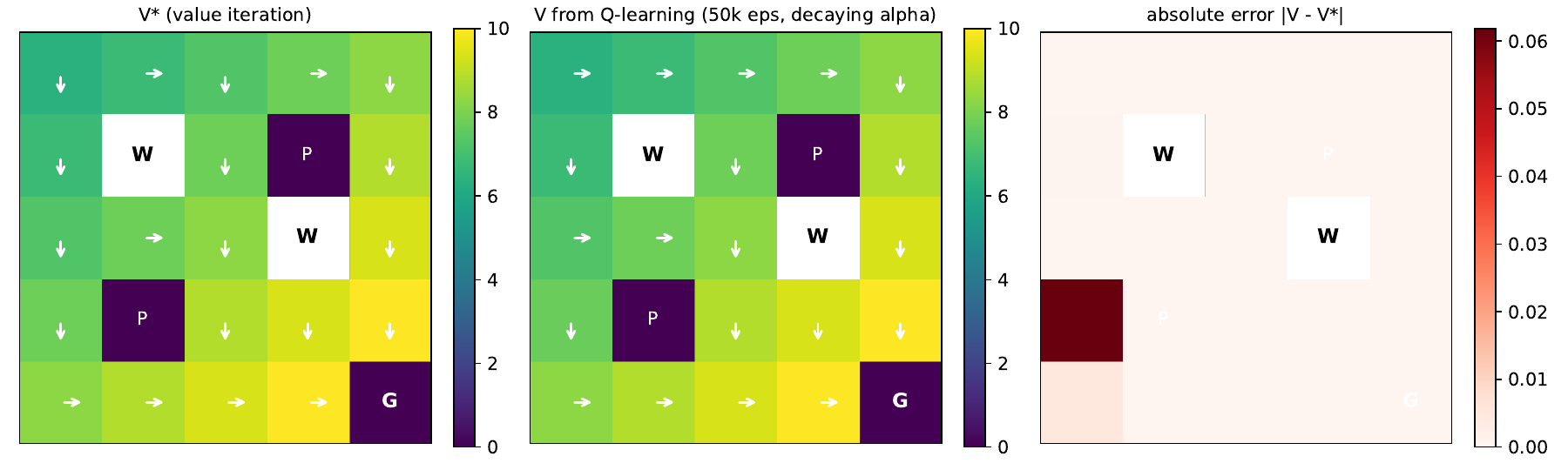}
\caption{Correctness of the tool's algorithm. Left: $V^\ast$ and optimal policy from value iteration. Center: value function and greedy policy learned by the tool's Q-learning rule (50k episodes, decaying $\alpha$, seed 0). Right: absolute error, concentrated in rarely-visited off-path states. W = wall, P = pit, G = goal.}
\label{fig:correctness}
\end{figure}

Across 10 seeds, the greedy policy matches an optimal action (Eq.~\ref{eq:agreement}) in $98.5\%$ of non-terminal states on average (minimum $95\%$, i.e.\ at most one state per run), and the maximum absolute error $\max_s |\max_a Q(s,a) - V^\ast(s)|$ averages $0.34$. Figure~\ref{fig:correctness} shows the learned value function converging to the ground truth. The residual disagreements are confined to states far from the start that the $\varepsilon$-greedy behavior policy rarely visits---itself a faithful illustration of a real property of Q-learning (off-path states receive few updates) rather than an implementation artifact, and one we now use as an advanced discussion point in the lesson.

\subsection{Hyperparameter sweeps: every lesson, reproduced}
\label{sec:sweeps}

\begin{figure}[t]
\centering
\includegraphics[width=\linewidth]{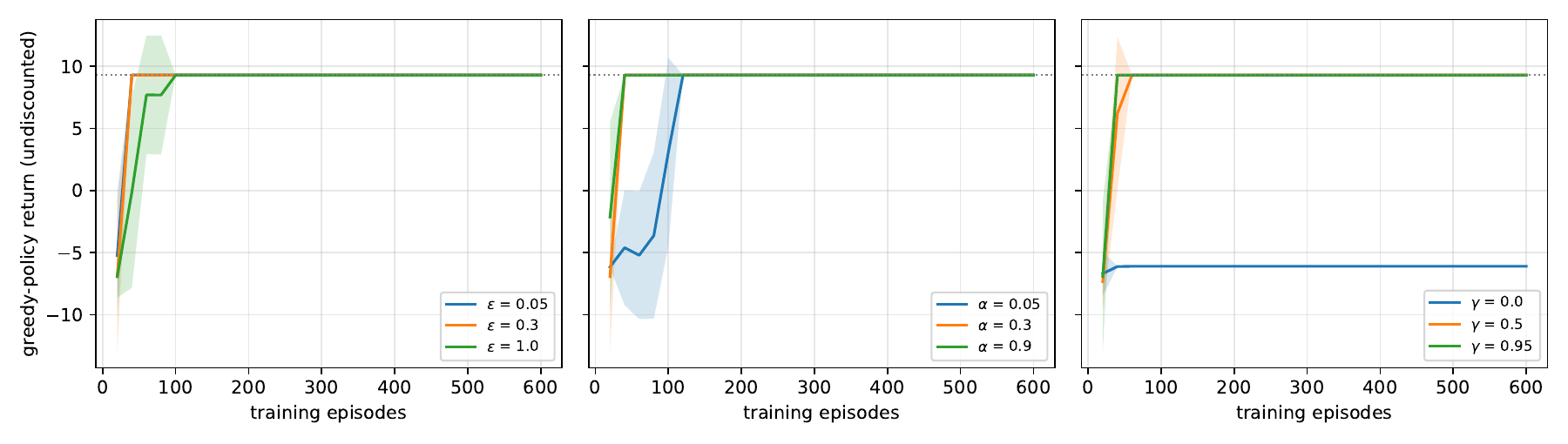}
\caption{Greedy-policy return during training for the sweeps in lesson moment (4): exploration $\varepsilon$ (left), learning rate $\alpha$ (center), discount $\gamma$ (right). Mean $\pm$ 1 s.d.\ over 10 seeds; dotted line marks the optimal return $9.3$. Note $\gamma=0$ never solves the task, and $\varepsilon=1$ still converges---Q-learning is off-policy.}
\label{fig:sweeps}
\end{figure}

Each claim in lesson moment (4) is tested as an experiment: for each setting we train for 600 episodes and evaluate the greedy policy every 20 episodes by its undiscounted return from the start state (optimal $= 9.3$; runs whose greedy policy loops are scored as failures). Figure~\ref{fig:sweeps} reports means $\pm$ one standard deviation over 10 seeds.

The results confirm the tool's narrative quantitatively. All $\varepsilon\in\{0.05, 0.3, 1.0\}$ eventually reach the optimal return---including $\varepsilon=1$, a nice classroom demonstration that Q-learning is off-policy and learns the greedy policy even from purely random behavior, though with slower, noisier progress. All $\alpha\in\{0.05,0.3,0.9\}$ converge on this small problem, with $\alpha=0.05$ visibly slower and $\alpha=0.9$ noisier in early training. The discount sweep is the starkest: $\gamma\in\{0.5,0.95\}$ reach the optimum, while $\gamma=0$ never solves the task (final return $-6.1$)---the myopic agent cannot propagate the goal reward backwards, exactly the ``short-sighted agent'' the lesson describes.

\subsection{Editing rewards: two distinct failure modes}
\label{sec:misspec}

Raising the pit reward causes the agent to end in a pit---but \emph{why} it does so depends on the reward value in a way that is easy to misread, and that distinction is the real lesson. We therefore treat two regimes separately, and in each we compare the learned greedy policy against the value-iteration optimal policy on the \emph{same} edited MDP. The optimal policy prefers a pit only once the pit reward exceeds a threshold, which we locate by bisection at $R_{\text{pit}}^{\dagger}=7.774$ (for the default step penalty and $\gamma=0.95$): below it, the shorter four-step path to a pit is worth less than the longer eight-step path to the goal; above it, the pit becomes genuinely optimal. Figure~\ref{fig:misspec} shows both regimes.

\subsubsection*{(a) $R_{\text{pit}}=+5$: exploration failure, \emph{not} reward hacking}

Setting the pit reward to $+5$ is the natural first thing a student tries, and the agent duly ends in a pit in $10/10$ seeds, with $100\%$ of training episodes terminating in a pit. It is tempting---and we caution instructors against this reading---to call this reward hacking. It is not. Value iteration on this exact MDP gives $V^\ast(s_0)=6.38$ with an optimal policy that still goes to the \emph{goal} (Figure~\ref{fig:misspec}, top-left): reaching a pit is worth only about $4.0$ in return, so pit-seeking is \emph{sub-optimal} under the very reward that was written. The agent fails not because it optimized a bad objective correctly, but because it never adequately optimized at all: the pit lies closer to the start and, being terminal, ends the episode before the agent can discover the goal, so the goal's higher value rarely propagates back. This is premature convergence to a local optimum driven by an exploration failure---a real and important RL phenomenon, but a different one from reward misspecification. The cross-check against the Bellman-optimal policy is exactly what separates the two, and we recommend it be performed in class rather than inferred from the agent's behavior alone.

\subsubsection*{(b) $R_{\text{pit}}=+20$: genuine reward misspecification}

\begin{figure}[t]
\centering
\includegraphics[width=0.86\linewidth]{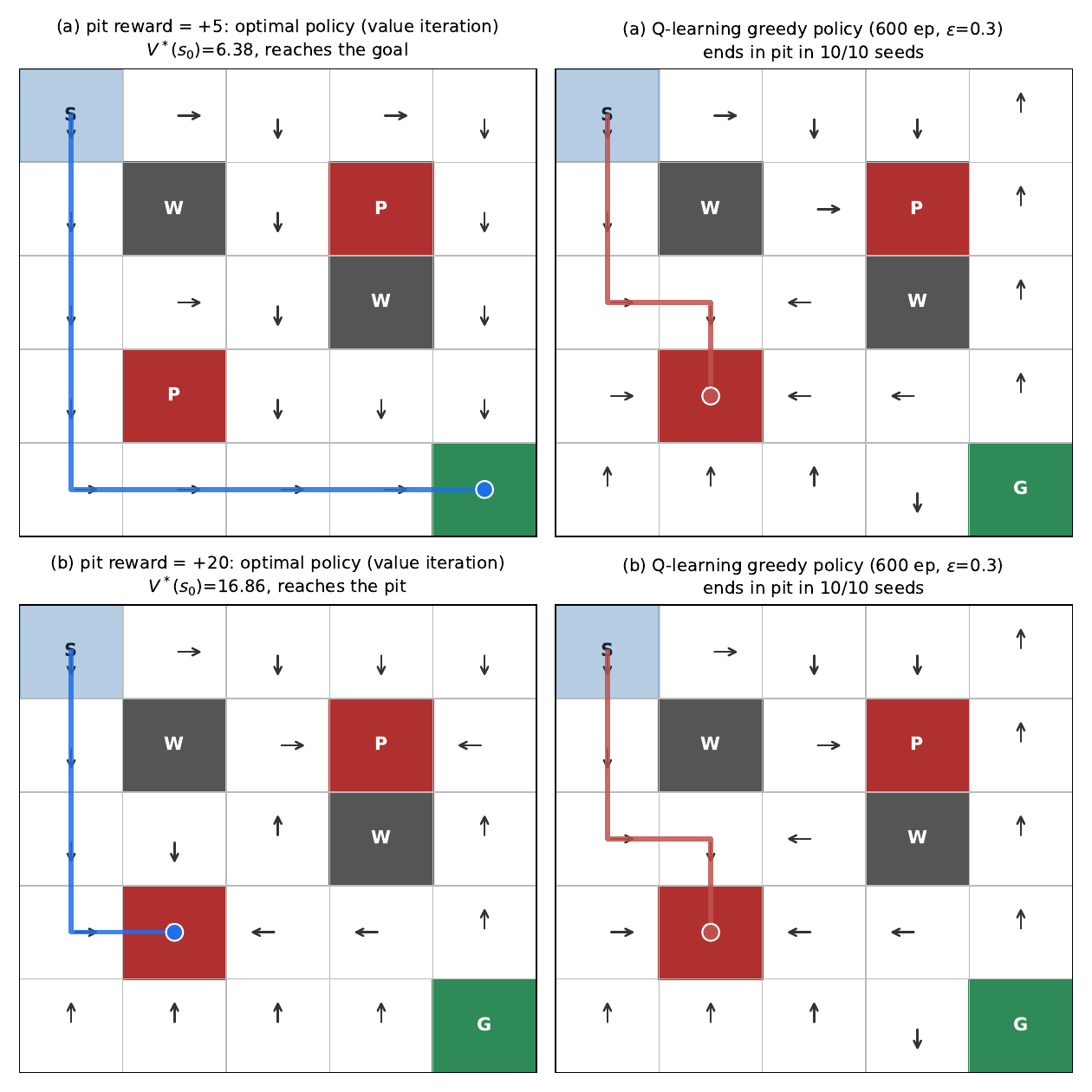}
\caption{Editing the pit reward produces two distinct failure modes, each shown as the Bellman-optimal policy (blue, value iteration) beside the learned greedy policy (red, Q-learning, 600 episodes, $\varepsilon=0.3$); the thick line traces the greedy path from the start. \textbf{Top ($R_{\text{pit}}=+5$):} the optimal policy still reaches the goal ($V^\ast(s_0)=6.38$), yet Q-learning converges to a pit in $10/10$ seeds---an \emph{exploration failure}, since pit-seeking is sub-optimal under the written reward. \textbf{Bottom ($R_{\text{pit}}=+20$, above the $7.774$ threshold):} the optimal policy genuinely prefers the pit ($V^\ast(s_0)=16.86$) and Q-learning agrees ($10/10$)---\emph{true reward misspecification}. Behaviorally identical, diagnostically opposite.}
\label{fig:misspec}
\end{figure}

To demonstrate reward hacking proper, the written reward must actually make the undesired behavior optimal. Setting the pit reward to $+20$ (above the $7.774$ threshold) does this: value iteration now yields $V^\ast(s_0)=16.86$ with an optimal policy that deliberately terminates in the nearest pit (Figure~\ref{fig:misspec}, bottom-left), and Q-learning recovers that same pit-seeking policy in $10/10$ seeds (bottom-right). Here the agent is behaving correctly with respect to a badly specified objective: it optimizes the reward as written, not as intended \cite{amodei2016}. This is the honest two-minute illustration of reward misspecification, and it is verifiable precisely because the learned policy agrees with the ground-truth optimal policy on the edited MDP.

\paragraph{The pedagogical payoff.} The contrast is more valuable than either case alone. The same one-line edit produces pit-seeking in both regimes, yet the cause is opposite---an under-optimized agent chasing a convenient shortcut ($+5$) versus a well-optimized agent serving a broken objective ($+20$). Distinguishing them requires computing the Bellman-optimal policy and comparing, which the tool's editable rewards and the accompanying reference implementation make a two-minute classroom exercise. That the two look identical from behavior but differ under analysis is, we think, a more durable lesson about diagnosing RL systems than a single tidy demonstration would have been---and it is a caution the authors themselves had to learn, having initially misread regime (a) as reward hacking.

\section{Discussion and Limitations}
\label{sec:limitations}

\paragraph{What the evaluation does and does not show.} Our evaluation establishes that the tool is algorithmically faithful and that its pedagogical demonstrations are reproducible properties of the underlying MDP, not fortunate animations. It does not establish learning gains in students; that requires a classroom study (pre/post concept inventory and usability instruments), which we have deliberately deferred and plan as the immediate next step. We regard technical validation as a prerequisite for such a study rather than a substitute: a tool whose claims are unverified should not be the treatment condition of an educational experiment.

\paragraph{Scope of the tool.} Q-Learning Lab teaches \emph{tabular} Q-learning on a deterministic, fully observable, single-agent gridworld. Stochastic transitions, function approximation, continuing tasks, and multi-agent settings are out of scope by design: the tool's value lies in exhaustively exposing one small algorithm, and every extension would trade transparency for coverage. The natural continuation---replacing the visible table with a small neural network while keeping the visible replay buffer---is left to a companion tool. A second continuation runs sideways rather than deeper: a student who has understood a single agent's Bellman update, and who has learned from Section~\ref{sec:misspec} to check behavior against a computed optimum, is equipped to ask what happens when the reward of one agent depends on the behavior of another. Modular multi-agent testbeds that keep dynamics, perception, and incentives separately inspectable \cite{atif2025} are a natural next environment for exactly that question, and we regard our tool as a preparation for such infrastructure rather than a competitor to it: the concepts the trace makes visible here (the update, the exploration decision, the reward's authorship of behavior) are precisely the concepts that become hard to see once several agents are learning at once.

\paragraph{Ceiling effects of a small MDP.} On a $5\times5$ deterministic grid, most hyperparameter settings eventually succeed (Section~\ref{sec:sweeps}), which compresses some contrasts (e.g., all tested $\alpha$ values converge). We consider this acceptable for a first encounter---failures like $\gamma=0$ and pit-seeking remain vivid---but instructors should not present the tool as evidence that hyperparameters ``do not matter.''

\paragraph{Trace size.} The on-screen log keeps the latest 500 steps and the fast-training mode skips logging for responsiveness; long training runs are therefore only partially traced. This is a deliberate trade-off in favor of interactivity, and the CSV export captures everything that was logged.

\section{Conclusion}

Q-Learning Lab reframes an RL teaching tool from an animation to be watched into an instrument that produces data to be analyzed. Its three distinguishing features---live numeric substitution into the Bellman update, decision-complete trace logging, and one-click export into a learn--export--analyze loop---are absent from existing gridworld visualizers, and our experiments verify that everything the tool demonstrates is a correct and reproducible property of the algorithm it implements. The tool is a single bilingual HTML file that runs anywhere a browser runs, accompanied by a 50-minute lesson plan and the full experiment suite of this paper.

\section*{Availability}

The tool, the Thai/English interface, the lesson plan, the Python reference implementation, and all experiment scripts and figures are available at \url{https://github.com/excel007/qlearning}. The tool is a single file (\texttt{qlearning\_lab.html}) and runs offline in any modern browser.


\begin{thebibliography}{9}
\itemsep 2pt

\bibitem[Amodei et al.(2016)]{amodei2016}
D.~Amodei, C.~Olah, J.~Steinhardt, P.~Christiano, J.~Schulman, and D.~Man\'e.
\newblock Concrete problems in AI safety.
\newblock \emph{arXiv preprint arXiv:1606.06565}, 2016.

\bibitem[Atif(2025)]{atif2025}
A.~Atif and contributors.
\newblock Predator--Prey Gridworld Environment: A deterministic modular testbed for multi-agent reinforcement learning.
\newblock \url{https://github.com/ProValarous/Predator-Prey-Archetype-Gridworld-Environment}, 2025.

\bibitem[Baker and Yacef(2009)]{baker2009}
R.~S. J.~d. Baker and K.~Yacef.
\newblock The state of educational data mining in 2009: A review and future visions.
\newblock \emph{Journal of Educational Data Mining}, 1(1):3--17, 2009.

\bibitem[DeNero and Klein(2010)]{denero2010}
J.~DeNero and D.~Klein.
\newblock Teaching introductory artificial intelligence with Pac-Man.
\newblock In \emph{Proceedings of the First AAAI Symposium on Educational Advances in Artificial Intelligence (EAAI)}, 2010.

\bibitem[Juliani(2025)]{juliani2025}
A.~Juliani.
\newblock The beginner's RL playground: A simple interactive website for grokking reinforcement learning.
\newblock \url{https://github.com/awjuliani/web-rl-playground}, 2025.

\bibitem[Karpathy(2020)]{karpathy2020}
A.~Karpathy.
\newblock micrograd: A tiny scalar-valued autograd engine and a neural net library on top of it.
\newblock \url{https://github.com/karpathy/micrograd}, 2020.

\bibitem[Kolb(1984)]{kolb1984}
D.~A. Kolb.
\newblock \emph{Experiential Learning: Experience as the Source of Learning and Development}.
\newblock Prentice-Hall, Englewood Cliffs, NJ, 1984.

\bibitem[Mnih et al.(2015)]{mnih2015}
V.~Mnih, K.~Kavukcuoglu, D.~Silver, et~al.
\newblock Human-level control through deep reinforcement learning.
\newblock \emph{Nature}, 518(7540):529--533, 2015.

\bibitem[Papert(1980)]{papert1980}
S.~Papert.
\newblock \emph{Mindstorms: Children, Computers, and Powerful Ideas}.
\newblock Basic Books, New York, 1980.

\bibitem[parasdahal(2018)]{tinyrl}
P.~Dahal.
\newblock tinyrl: Animated interactive visualization of RL algorithms.
\newblock \url{https://github.com/parasdahal/tinyrl}, 2018.

\bibitem[Rasoul(2024)]{rasoul}
Rasoul77.
\newblock gridworld-qlearning: An educational implementation of Q-learning in a 2D grid world.
\newblock \url{https://github.com/Rasoul77/gridworld-qlearning}, 2024.

\bibitem[Siemens(2013)]{siemens2013}
G.~Siemens.
\newblock Learning analytics: The emergence of a discipline.
\newblock \emph{American Behavioral Scientist}, 57(10):1380--1400, 2013.

\bibitem[Virtual Labs(2023)]{virtuallabs}
Virtual Labs, IIIT Hyderabad.
\newblock Q-learning experiment.
\newblock \url{https://virtual-labs.github.io/exp-q-learning-iiith/}, 2023.

\bibitem[Watkins and Dayan(1992)]{watkins1992}
C.~J. C.~H. Watkins and P.~Dayan.
\newblock Q-learning.
\newblock \emph{Machine Learning}, 8(3--4):279--292, 1992.

\bibitem[Winne(2010)]{winne2010}
P.~H. Winne.
\newblock Improving measurements of self-regulated learning.
\newblock \emph{Educational Psychologist}, 45(4):267--276, 2010.

\end{thebibliography}
\end{document}